\documentclass[12pt]{article}
\usepackage[utf8]{inputenc}
\usepackage{graphicx}
\usepackage{ulem}
\usepackage{appendix}
\usepackage{amssymb,amsmath,amsthm}
\usepackage{natbib}
\usepackage{comment}
\usepackage[margin=1in]{geometry}
\usepackage{booktabs}
\usepackage{xcolor}
\usepackage{url}
\usepackage{bbm}
\usepackage[width=0.86\textwidth]{caption}
\usepackage{authblk}

\DeclareMathOperator*{\argmin}{arg\,min}

\newcommand{\dnorm}{\mathcal{N}}
\newcommand{\bstau}{\boldsymbol \tau}

\newcommand{\hbstau}{\boldsymbol{\hat \tau}}
\newcommand{\tauu}{\boldsymbol{\hat \tau_u}}
\newcommand{\taub}{\boldsymbol{\hat \tau_b}}
\newcommand{\tauuk}{\hat \tau_{uk}}
\newcommand{\taubk}{\hat \tau_{bk}}
\newcommand{\tauw}{\boldsymbol{\hat \tau_w}}
\newcommand{\bst}{\boldsymbol \tau}
\newcommand{\bsSig}{\boldsymbol \Sigma}
\newcommand{\bsxi}{\boldsymbol \xi}

\newcommand{\htauo}{\boldsymbol{\hat \tau_b}}

\newtheorem{definition}{Definition}

\def\P{\mathbb{P}}
\def\E{\mathbb{E}}
\def\var{\text{var}}

\newcommand{\diag}{\mathrm{diag}}
\newcommand{\Tr}{\text{tr}}

\newcommand{\hbsSig}{\boldsymbol{\hat \Sigma}}
\newcommand{\ident}{\boldsymbol I}
\newcommand{\bsdelt}{\boldsymbol \delta}

\newcommand{\bskap}{ \boldsymbol\kappa}

\newcommand{\tran}{\mathsf{T}}
\newcommand{\bsp}{\boldsymbol{ \hat \psi}}

\newcommand{\bspmm}{\boldsymbol{\hat \psi_{\textbf{mm}}}}
\newcommand{\bspmmone}{\boldsymbol{\hat \psi_{\textbf{mm,1}}}}
\newcommand{\bspmmtwo}{\boldsymbol{\hat \psi_{\textbf{mm,2}}}}
\newcommand{\bspmle}{\boldsymbol{\hat \psi_{\textbf{mle}}}}
\newcommand{\bspure}{\boldsymbol{\hat \psi_{\textbf{ure}}}}

\newcommand{\gammamm}{\hat \gamma_{\text{mm}}^2}
\newcommand{\etamm}{\hat \eta_{\text{mm}}^2}

\newcommand{\gammammone}{\hat \gamma_{\text{mm,1}}^2}
\newcommand{\etammone}{\hat \eta_{\text{mm,1}}^2}

\newcommand{\gammammtwo}{\hat \gamma_{\text{mm,2}}^2}

\newcommand{\gammamle}{\hat \gamma_{\text{mle}}^2}
\newcommand{\etamle}{\hat \eta_{\text{mle}}^2}

\newcommand{\ure}{\text{URE}}
\newcommand{\gammaure}{\hat \gamma_{\text{ure}}^2}
\newcommand{\etaure}{\hat \eta_{\text{ure}}^2}

\title{Empirical Bayes Double Shrinkage for Combining Biased and Unbiased Causal Estimates}
\author[1]{Evan T. R. Rosenman}
\author[2]{Francesca Dominici}
\author[3]{Luke Miratrix}
\affil[1]{Department of Mathematical Sciences, Claremont McKenna College}
\affil[2]{Department of Biostatistics, Harvard T.H. Chan School of Public Health}
\affil[3]{Harvard Graduate School of Education}

\begin{document}

\maketitle

\abstract{Motivated by the proliferation of observational datasets and the need to integrate non-randomized evidence with randomized controlled trials, causal inference researchers have recently proposed several new methodologies for combining biased and unbiased estimators. We contribute to this growing literature by developing a new class of estimators for the data-combination problem: double-shrinkage estimators. Double-shrinkers first compute a data-driven convex combination of the the biased and unbiased estimators, and then apply a final, Stein-like shrinkage toward zero. Such estimators do not require hyperparameter tuning, and are targeted at multidimensional causal estimands, such as vectors of conditional average treatment effects (CATEs). We derive several workable versions of double-shrinkage estimators and propose a method for constructing valid Empirical Bayes confidence intervals. We also demonstrate the  utility of our estimators using simulations on data from the Women's Health Initiative. }

\tableofcontents

\section{Introduction}

The modern proliferation of observational datasets – in applications such as disease surveillance, voter mobilization, and e-commerce – provides an opportunity to improve causal estimation. These data are typically large, inexpensive to obtain, and representative of target populations. However, treatments are not randomized in observational studies, meaning that treated and control units may differ in important ways. Hence, there is a fundamental challenge to the utility of observational studies: these studies frequently suffer from unmeasured confounding. As a result, even with reasoned statistical adjustments, the causal estimates obtained from observational data are often biased. 

By contrast, the virtues of randomized controlled trials (RCTs) are widely known. Experimental data yield unbiased estimates of causal effects under very mild  assumptions. However, high-quality randomized trials are expensive to conduct, and tend to have a limited number of participants. They are frequently underpowered for the estimation of causal effects on subgroups of the population. Given these challenges using experimental data alone, one can find a chorus of recent papers \citep[e.g.][]{shalit2020can, mueller2018methods} advocating for methodological advancements in combining experimental and observational data to obtain more statistically efficient estimates of causal effects. Many researchers have heeded this call in the past few years \citep{kallus2018removing, cheng2021adaptive, yang2020elastic, colnet2020causal}.

This papers adds to the growing literature around data-combination approaches by proposing a new class of estimators for combining biased and unbiased estimators: ``double shrinkers." These estimators work by first computing data-driven weights to apply to the biased and unbiased estimators, yielding a convex combination of the estimators. Then, double-shrinkers apply a final, Stein-like shrinkage toward zero. The latter step can be viewed as a form of regularization, inducing some additional bias but reducing the variance of the estimator.

We operate under squared error loss, meaning we seek to obtain estimates $\hbstau \in \mathbb{R}^K$ of causal effects $\bstau$ such that 
\[ \mathcal{L}(\hbstau, \bstau) = \sum_{k = 1}^K (\hat \tau_k - \tau_k)^2\] 
is typically small. Under this loss, the ``dual shrinkage" property turns out to generate significantly improved estimation performance in some settings. Moreover, constructing the estimator from an explicit generative model allows for straightforward construction of confidence intervals with robust coverage properties. 


\subsection{Prior Work}

This manuscript builds primarily upon four previous works:   \cite{green2005improved}, \cite{rosenman2020combining}. \cite{green1991james}, and \cite{xie2012sure}.  

\cite{green2005improved} considers how to combine biased and unbiased estimators in the Empirical Bayes framework. The authors suppose they have access to two $K$-dimensional estimators, $\tauu$ and $\taub$, such that $\tauu$ has mean $\bst$ and $\taub$ has mean $\bst - \bsxi$, where $\bsxi$ represents a $K$-dimensional bias vector. The estimand is $\bst$, so $\tauu$ is an unbiased estimator while $\taub$ is a biased estimator. The unbiased estimator is assumed normally distributed and heteroscedastic, such that $\var(\tauu) = \bsSig_u = \diag( \{\sigma_{rk}^2\}_{k = 1}^K )$. No other assumptions are placed on the distribution of $\taub$. 

The authors derive two estimators heuristically, and suggest them for use in different contexts. The first estimator is intended to perform well under the precision-weighted squared-error loss,
\[ \mathcal{L}(\hbstau, \bst) = \sum_{k = 1}^K \frac{\left( \hat \tau_k - \tau_k \right)^2}{\sigma_k^2} \,.\] 
Under this loss, the estimator 
\[ \bsdelt_1 = \taub + \left( 1 - \frac{a}{(\tauu - \tauu)^\tran  \hbsSig_u^{-1} \left( \tauu - \taub \right)}\right) \left(\tauu - \taub \right) \] 
dominates $\tauu$ (in the decision theoretic sense), as long as $0 < a < 2(K-2)$ and $\hbsSig_u$ is correctly estimated. 

The authors propose a different estimator for the more conventional squared error loss, 
\[ \bsdelt_2 = \taub + \left(\ident_K - \frac{a \hbsSig_u^{-1}}{(\tauu - \taub)^\tran  \hbsSig_u^{-2} (\tauu - \taub)} \right)(\tauu - \taub )\,. \] 
This estimator can also be shown to dominate $\tauu$ if $0 < a < 2(K-2)$. The authors argue that the shrinkage parameter $a$ should be selected as $K-2$ for both estimators. 

\cite{rosenman2020combining} adopted the same setting as \cite{green2005improved}, and built upon its results by introducing an alternative, non-heuristic method for deriving data-combination shrinkage estimators. The paper proposes constructing combinations of biased and unbiased estimators by appealing to an unbiased risk estimate (URE) analogous to the classical Stein's unbiased risk estimate \citep[SURE;][]{stein1981estimation}. The proposed process involves two steps: positing a shrinkage structure, and then deriving the values of tuning parameters by minimizing over the URE. The latter idea -- minimizing an unbiased estimate of risk to obtain the value of hyperparameters -- has substantial history in statistics \citep{xie2012sure}. Using this approach, two new estimators are developed in \cite{rosenman2020combining}: $\bskap_{1+}$, which shrinks all components of $\tauu$ by the same multiplicative factor toward $\taub$; and $\bskap_{2+}$, which shrinks each component proportionally to the variance of each entry of $\tauu$. These estimators are competitive with $\bsdelt_1$ and $\bsdelt_2$ in simulation and in a real data analysis using data from the Women's Health Initiative. 

A relevant precedent to \cite{green2005improved} is \cite{green1991james}, which considered the same problem but in the case of homoscedastic $\tauu$ and $\taub$. The earlier paper proposes a simpler estimator, 
\[ \bsdelt = \taub + \left( 1 - \frac{(K-2)\sigma^2}{||\taub - \tauu ||^2} \right)_+ (\tauu - \taub)\,, \]
where $\sigma^2 \in \mathbb{R}$ is the variance of each entry in $\tauu$. $\bsdelt$ is not derived explicitly in \cite{green1991james}, but the authors mention in passing that their estimator can be constructed from a hierarchical model of the data-generating process, namely 
\begin{equation}\label{eq:hierModel}
\begin{aligned}
p(\bst) & \propto c, \\
\bsxi & \sim N(0, \gamma^2 \ident_K),  \\
\tauu \mid \bst &\sim N(\bst, \sigma^2 \ident_k), \text{ and } \\
\taub \mid \bst, \bsxi &\sim N(\bst + \bsxi, \phi^2 \ident_K),
\end{aligned}
\end{equation}
where the first line represents a noninformative (i.e. improper locally uniform) prior on $\bst$.
 
Lastly, \cite{xie2012sure} considers a different setting: that of the classical James-Stein estimator \citep{stein1956inadmissibility}, in which the goal is to shrink a multivariate normal mean vector toward a central point. A key complication in \cite{xie2012sure} is that the authors do not impose a homoscedasticity assumption on the multivariate normal, unlike Stein in the original work. The literature contains no consensus estimator in the heteroscedastic case. In  \cite{xie2012sure}, as in \cite{green1991james}, the authors use a hierarchical model to derive a functional form for a shrinkage estimator. They then propose three different heuristics -- based on moment-matching, maximum likelihood estimation, and SURE minimization -- to construct usable estimators. In their setting, they find that the estimator derived from SURE minimization often performs best. 

Beyond the direct antecedents to this paper, we note that there are several alternate perspectives on estimator construction for the data-combination problem. 
A growing literature on this problem has developed in causal inference. 
Several papers propose methods to correct the bias in the observational study estimates using the joint information from the RCT and the observational study. \cite{kallus2018removing} proposes a deconfounding technique that relies on estimating a correction term, under the assumption that the confounding bias has a parametric structure that can be modeled and extrapolated to other parts of the covariate space. 
More recently, \cite{yang2020improved} considers a ``confounding function" that captures the bias due to unmeasured confounding at each covariate value. They assume the conditional average treatment effect function and the confounding function both have parametric forms, and derive a semiparametric efficient estimator for the model parameters for both functions. \cite{colnet2020causal} provides an excellent overview of these and other proposed methods. 

Several recent papers have also proposed adaptive schemes for trading off between observational and experimental estimators. \cite{cheng2021adaptive} proposes an approach that approximates the optimal linear combination of the observational and experimental estimators. Their estimator weights heavily toward the RCT estimator when bias is detected in the observational study, but pools the two data sources when bias is negligible. \cite{yang2020elastic} adopts a testing-based approach, wherein the equality of means between the observational and experimental estimators is tested and the data is pooled only if the test fails to reject the null. \cite{chen2021minimax} proposes a soft-thresholding estimator for combining the data sources, and also demonstrates that such an estimator achieves the minimax convergence rate for mean squared error of the true causal effects, up to poly-log factors. Lastly, \cite{oberst2022bias} considers the problem in the case where the parameter of interest is unidimensional, and propose a simple data-combination estimator with a provably bounded bias. 

\subsection{Contributions}

Recent advances offer practitioners a menu of options for combining data from observational and experimental settings. Yet the choice of estimator for any specific task is not immediately clear. Even in the univariate case, \cite{oberst2022bias} show that different estimators do better when the bias in $\taub$ is lower or higher -- but the magnitude of the bias is typically not known to the researcher a priori. Moreover, some methods require a choice of hyperparameters or significance thresholds (such as those of \cite{cheng2021adaptive}, \cite{chen2021minimax}, and \cite{yang2020elastic}) while others require untestable assumptions (such as that of \cite{kallus2018removing}). Lastly, not all methods admit confidence intervals or clearly establish the assumptions required for coverage guarantees. 

We propose an approach for the case where the unbiased estimator $\tauu$ and $\taub$ are $K$-dimensional and heteroscedastic. Our work complements the existing literature in several ways. First, we do not require hyperparameter tuning for estimation or inference, nor do we impose any assumptions beyond the normality of $\tauu$ and $\taub$. Second, our estimator arises from a straightforward derivation using a hierarchical model of the data-generating process. This approach is intuitively appealing; extends prior work from \cite{green1991james}; and allows for straightforward construction of  Empirical Bayes confidence intervals with robust coverage properties. Third, our estimator is novel in its functional form, utilizing not only a data-driven weighting scheme to trade off between $\tauu$ and $\taub$, but also a Stein-like shrinkage toward a central point. This ``dual shrinkage" property generates significantly improved performance in simulations.

The remainder of this paper proceeds as follows. 
Section \ref{sec:results} introduces our notation and assumptions; works through the construction of the different versions of our estimator; and also details the construction of valid Empirical Bayes confidence intervals Section \ref{sec:whi} contains a detailed simulation study using data from the Women's Health Initiative, a 1991 study of the effect of hormone therapy on health outcomes for postmenopausal women. We consider both estimation and inference tasks, finding that the MLE-based double shrinker is particularly performant versus competitor estimators. Section \ref{sec:conclusion} concludes.

\section{Main Results}\label{sec:results}

\subsection{Hierarchical Model}\label{sec:hierModel}

We adopt the same notation as in the prior section, denoting the unbiased estimator as $\tauu$, the biased estimator as $\taub$, the true causal effects as $\bstau$, and the vector of biases in $\taub$ as $\bsxi$. All four objects lie in $\mathbb{R}^K$. We assume both $\tauu$ and $\taub$ are heteroscedastic. Moreover, in keeping with the traditional approach in Empirical Bayes, we assume their diagonal covariance matrices $\bsSig_u = \diag(\sigma_{uk}^2) \in \mathbb{R}^{K \times K}$ and $\bsSig_b = \diag(\sigma_{bk}^2) \in \mathbb{R}^{K \times K}$ are known. Our goal is to obtain an estimator of $\bstau$ that performs well under squared error loss.


Rather than deriving our estimator by positing its functional form and optimizing over Stein's unbiased risk estimate  \citep[as in][]{rosenman2020combining}, we instead obtain it by proposing a hierarchical model for the data-generating process. We assume the following  model for the data,

\begin{equation}\label{eq:obsData}
\begin{aligned}
\bstau &\sim \mathcal{N} \left(0, \eta^2 \ident_K \right), \\
\bsxi & \sim \mathcal{N} \left(0, \gamma^2 \ident_K \right), \\
\tauu \mid \bstau & \sim \mathcal{N} \left( \bstau, \bsSig_u \right), \text{ and } \\
\taub \mid \bstau, \bsxi &\sim \mathcal{N} \left( \bstau + \bsxi, \bsSig_b \right) ,
\end{aligned}
\end{equation}
where $\eta^2$ and $\gamma^2$ are unknown hyperparameters. This proposed data-generating process is similar to the model (\ref{eq:hierModel}) from  \cite{green1991james}, but it differs in a few key ways, as described below. 


The model described in (\ref{eq:obsData}) imposes a normal prior on both $\bstau$ and $\bsxi$, with both centered at 0. Conditional on a draw of these parameters, $\tauu$ is  normally distributed about $\bstau$ with diagonal covariance matrix $\bsSig_u$, while $\taub$ is assumed normally distributed about $\bstau + \bsxi$ with  covariance matrix $\bsSig_b$. 

The sampling distributions of $\tauu$ and $\taub$ are themselves an assumption. We are supposing that a Central Limit Theorem holds for each estimator; that we have boundedness and sufficient sample size such that the CLT can appropriately describe the sampling distribution; and that our estimation technique is independent within each stratum, such that the covariance matrices of both estimators are diagonal. 

We do not put priors on $\bsSig_u = \diag\left(\sigma_{u1}^2, \dots, \sigma_{uK}^2 \right)$ and $\bsSig_b = \diag \left(\sigma_{b1}^2, \dots, \sigma_{bK}^2 \right)$, as these are assumed known (though in practice they are estimated from the data). The given model differs from the one in \cite{green1991james} in that we allow for heteroscedasticity of $\tauu$ and $\taub$, and we use a normal prior rather than an improper uniform prior for $\bstau$.

Because the priors and likelihoods are normally distributed, we have conjugacy such that the posterior distribution of $\bstau$ and $\bsxi$ will be jointly normal. We are interested in the posterior mean of $\bstau$, which evaluates to 
\begin{equation}\label{eq:postMean}
\begin{aligned}
\E \left( \bstau \mid \tauu, \taub \right) &= \int_{-\infty}^{\infty} \E \left( \bstau, \bsxi \mid \tauu, \taub \right) \\
& =  \left\{ \frac{\eta ^2 \left(\hat \tau_{uk} \left(\gamma ^2+\sigma_{bk}^2\right)+\sigma_{uk}^2
   \hat \tau_{bk} \right)}{\gamma ^2 \left(\eta ^2+\sigma_{uk}^2\right)+\eta ^2
   \left(\sigma_{uk}^2+\sigma_{bk}^2\right)+\sigma_{uk}^2 \sigma_{bk}^2} \right\}_{k = 1}^K\,
\end{aligned}
\end{equation}
where $\hat \tau_{uk}$ and $\hat \tau_{bk}$ are the $k^{th}$ entries of $\tauu$ and $\taub$, respectively. 

This can be reorganized into a more suggestive form. We denote our shrinker as $\bsp(\gamma^2, \eta^2)$, a function of $\gamma^2$ and $\eta^2$. The $k^{th}$ entry of the shrinker can be written as: 
\begin{equation}\label{eq:shrinkForm}
\begin{aligned}
 \psi_k(\gamma^2, \eta^2) &=  \underbrace{\left( \frac{\eta^2 \left( \gamma^2 + \sigma_{bk}^2 + \sigma_{uk}^2 \right) }{\sigma_{uk}^2 \left( \gamma^2 + \sigma_{bk}^2 \right) + \eta^2 \left( \gamma^2 + \sigma_{bk}^2 + \sigma_{uk}^2 \right) } \right)}_{a_k} \cdot \left( \underbrace{ \frac{  \left(\gamma ^2+\sigma_{bk}^2\right)}{\gamma^2 + \sigma_{bk}^2 + \sigma_{uk}^2}}_{\lambda_k} \hat \tau_{uk}+ \underbrace{\frac{\sigma_{uk}^2}{\gamma^2 + \sigma_{bk}^2 + \sigma_{uk}^2}}_{1 - \lambda_k} \hat \tau_{bk} \right)
\end{aligned}
\end{equation}

In the form of (\ref{eq:shrinkForm}), we can interpret this posterior mean as ``doubly shrunken." The first term, denoted $a_k$, is a general shrinkage toward zero. This term is analogous to the shrinkage we observe when using the James-Stein estimator. If we take $\eta^2$ to infinity (i.e., make the prior on $\bstau$ a flat prior), then this term goes to $1$. 

The second term is a convex  combination of $\hat \tau_{uk}$ and $\hat \tau_{bk}$, with weights $\lambda_k$ and $1 - \lambda_k$. Observe that the weights are inversely proportional to the expected mean squared error of each estimator. If we take $\sigma_{uk}^2$ to 0, then the weight concentrates on $\hat \tau_{uk}$, and the opposite happens if we take $\gamma^2 + \sigma_{bk}^2$ to 0. 

\subsection{Operationalizing the Estimator}\label{subsec:op}
We cannot directly use (\ref{eq:shrinkForm}) as an estimator, because $\gamma$ and $\eta$ are unknown (indeed, they are just constructed from an imagined prior). Nonetheless, if we can obtain estimates $\hat \gamma^2$ and $\hat \eta^2$ of these parameters from the data, we can then define the functions $\boldsymbol a: \mathbb{R}^2 \to \mathbb{R}^K$ and $\boldsymbol \lambda: \mathbb{R}^2 \to \mathbb{R}^K$ according to 
\[ \boldsymbol a \left( \hat \gamma^2, \hat \eta^2 \right) = \bigg\{  a_k \left( \hat \gamma^2, \hat \eta^2 \right) \bigg\}_{k = 1}^K = \left\{ \left(  \frac{ \hat  \eta^2 \left(  \hat \gamma^2 + \sigma_{bk}^2 + \sigma_{uk}^2 \right)}{\sigma_{uk}^2 \left( \hat \gamma^2 + \sigma_{bk}^2 \right) + \hat  \eta^2 \left( \hat  \gamma^2 + \sigma_{bk}^2 + \sigma_{uk}^2 \right) } \right) \right\}_{k = 1}^K, \] 
and 
\[ \boldsymbol \lambda (\hat \gamma^2, \hat \eta^2) = \bigg\{  \lambda_k \left( \hat \gamma^2, \hat \eta^2 \right) \bigg\}_{k = 1}^K = \left\{ \left( \frac{   \hat \gamma ^2+\sigma_{bk}^2 }{\hat \gamma^2 + \sigma_{bk}^2 + \sigma_{uk}^2}\right) \right\}_{k = 1}^K. \] 
Lastly, we can define a usable shrinkage estimator as 
\[ \bsp(\hat \gamma^2, \hat \eta^2) = \boldsymbol a \left( \hat \gamma^2, \hat \eta^2 \right) \circ  \bigg( \boldsymbol \lambda (\hat \gamma^2, \hat \eta^2)  \circ \tauu + \big(\boldsymbol 1 - \boldsymbol \lambda (\hat \gamma^2, \hat \eta^2) \big) \circ  \taub \bigg) \]
where $\circ$ represents an elementwise (Hadamard) product,  

The remaining question is how precisely to estimate $\hat \gamma^2$ and $\hat \eta^2$. In analogy with \cite{xie2012sure}, we offer three possible paradigms: moment matching, Empirical Bayes maximum likelihood, and URE minimization. Each of these estimation procedures leads to a different version of our estimator. 

\subsubsection{Moment Matching}

One common approach is to find observable quantities whose expectations are equal to $\gamma^2$ and $\eta^2$ under hierarchical model (\ref{eq:obsData}). This moment-matching approach generates two candidate estimators. 

First, under model (\ref{eq:obsData}), observe that 
\begin{align*}
\E \left( || \taub - \tauu ||_2^2 \right) &=   \Tr(\bsSig_b) + \Tr(\bsSig_u) + K \gamma^2\,, \\
\E \left( || \taub ||_2^2 - || \tauu ||_2^2 \right) &=  \Tr(\bsSig_b) - \Tr(\bsSig_u) + K \gamma^2\,.
\end{align*}
Assuming $\bsSig_u$ and $\bsSig_b$ are estimated well in practice, we can estimate $\gamma^2$ using either
\[ \gammammone = \frac{1}{K} \left( || \tauu - \taub ||_2^2 - \Tr(\bsSig_u) - \Tr(\bsSig_b) \right)_+, \] 
or
\[ \gammammtwo = \frac{1}{K} \left( || \taub||_2^2 - || \tauu ||_2^2 + \Tr(\bsSig_u) - \Tr(\bsSig_b) \right)_+, \] 
where $u_+ = \max(u, 0)$ represents the positive-part estimator. 

Analogously we observe 
\[ \E \left( ||\tauu||_2^2 \right) = \Tr(\bsSig_u) + K \eta^2 \] 
and hence we can use the moment matching estimator 
\[ \etamm= \frac{1}{K} \left( ||\tauu||_2^2 - \Tr(\bsSig_u) \right)_+. \] 
Plugging in our two candidate moment-matching estimators for $\gamma^2$, and our single candidate moment-matching estimator for $\eta^2$, we arrive at two candidate moment-matching versions of our shrinkage estimator, 
\[ \bspmmone \equiv \bsp(\gammammone, \etamm) \hspace{5mm} \text{ and } \hspace{5mm}  \bspmmtwo \equiv \bsp(\gammammtwo, \etamm) . \]


\subsubsection{Empirical Bayes Maximum Likelihood}

An alternative approach is to use maximum likelihood. Under model (\ref{eq:obsData}), the marginal distributions of $\tauu$ and $\taub$ satisfy 
\[ \tauu \sim \mathcal{N} \left(0, \eta^2\ident_K + \bsSig_u \right) \hspace{5mm} \text{ and } \hspace{5mm} \taub \sim \mathcal{N} \left(0, \eta^2\ident_K + \gamma^2\ident_K + \bsSig_b \right)\,. \]
Hence, we can write 
\[ f(\tauu, \taub) \propto \prod_k \left(\eta^2 + \sigma_{uk}^2 \right)^{-1/2} e^{-\frac{\tauuk^2}{2\left(\eta^2 + \sigma_{uk}^2 \right)}}\times \prod_k  \left(\eta^2 + \gamma^2 + \sigma_{bk}^2 \right)^{-1/2} e^{-\frac{\taubk^2}{2\left(\eta^2 + \gamma^2 + \sigma_{bk}^2 \right)}} \,.\] 

Taking the logarithm, we obtain 
\[ -\frac{1}{2} \sum_k\left(  \log \left(\eta^2 + \sigma_{uk}^2 \right)  + \frac{\tauuk^2}{\left(\eta^2 + \sigma_{uk}^2 \right)} \right) + \left(\log \left(\eta^2 + \gamma^2 + \sigma_{bk}^2 \right) + \frac{\taubk^2}{\left(\eta^2 + \gamma^2 + \sigma_{bk}^2 \right)} \right) \,.\] 

This expression is concave, so we can obtain estimates $\etamle$ and $\gammamle$ as zeroes of its gradient, i.e. as solutions to the equations 
\begin{equation}\label{etaEq}
-\frac{1}{2}\sum_k \left( \frac{1}{\eta^2 + \sigma_{uk}^2} - \frac{\tauuk^2}{\left(\eta^2 + \sigma_{uk}^2 \right)^2} \right) +  \left( \frac{1}{\eta^2 + \gamma^2 + \sigma_{bk}^2} - \frac{\taubk^2}{\left(\eta^2 + \gamma^2 + \sigma_{bk}^2 \right)^2} \right)= 0 \,,
\end{equation}
and 
\begin{equation}\label{gammaEq}
-\frac{1}{2}\sum_k  \left( \frac{1}{\eta^2 + \gamma^2 + \sigma_{bk}^2} - \frac{\taubk^2}{\left(\eta^2 + \gamma^2 + \sigma_{bk}^2 \right)^2} -\right)  = 0 \,,
\end{equation}
where $\etamle = 0$ in the case where the system has no positive solution for $\eta$ and $\gammamle = 0$ when the system has no positive solution for $\gamma$.  Plugging in these estimates gives us the second version of our estimator,
\[ \bspmle \equiv \bsp(\gammamle, \etamle). \]




\subsubsection{Unbiased Risk Estimate Minimization}

A final approach eschews direct estimation of $\eta^2$ and $\gamma^2$, instead choosing their values to minimize an unbiased estimate of the shrinker's statistical risk. Using results from \cite{rosenman2020combining}, we can compute the risk of a shrinkage estimator $\bsp(\hat \gamma^2, \hat \eta^2)$. The full computation is given in Appendix \ref{app:ureComp}. The resultant risk value is 
\begin{equation}\label{eq:ureForm}
\begin{aligned}
\E \left( || \bsp(\hat \gamma^2, \hat \eta^2) - \bst ||_2^2 \mid \bst, \bsxi \right) = \hspace{1mm}&  \Tr ( \bsSig_u ) + \E \left( || \bsp(\hat \gamma^2, \hat \eta^2) - \tauu ||_2^2 \mid \bst, \bsxi \right) - \\ & 2\sum_k \E \bigg( \sigma_{uk}^2 \cdot \big(1 - a_k \left( \hat \gamma^2, \hat \eta^2 \right) \cdot \lambda_k \left( \hat \gamma^2, \hat \eta^2 \right) \big) \mid \bst, \bsxi \bigg)\,.
\end{aligned}
\end{equation}

Removing the (conditional) expectations, this yields an unbiased estimate of the statistical risk of shrinker $\bsp(\hat \gamma^2, \hat \eta^2)$, which can be computed directly from the data, i.e. 
\begin{equation}\label{ureDef}
\ure(\hat \gamma^2, \hat \eta^2) =  \Tr ( \bsSig_u ) +  || \bsp(\hat \gamma^2, \hat \eta^2) - \tauu ||_2^2   -  2\sum_k  \sigma_{uk}^2 \cdot \big(1 - a_k \left( \hat \gamma^2, \hat \eta^2 \right) \cdot \lambda_k \left( \hat \gamma^2, \hat \eta^2 \right) \big)   \,.
\end{equation}
Equation \ref{ureDef} is  useful, because it allows us to optimize hyperparameters over an unbiased estimate of the risk for the purposes of estimator design. Such an approach has substantial precedent in the literature \citep{li1985stein, xie2012sure}. We obtain 
\[ \bspure \equiv \bsp(\gammaure, \etaure), \hspace{5mm} \text{ where } \hspace{5mm} (\gammaure, \etaure) = \argmin_{\gamma^2 \geq 0, \eta^2 \geq 0} \ure( \gamma^2,  \eta^2)\,. \] 

\subsection{Relationships Among Estimators}

Curiously, our setting departs from that of \cite{xie2012sure} in an unexpected way. Xie and co-authors found that, when the multivariate Gaussian distribution of interest was homoscedastic, their three estimator versions -- moment-matching, maximum likelihood, and SURE-based -- coincided.  The different paradigms could therefore be seen as alternative approaches to deal with non-constant variance across components. 

In our context, the estimators do not all coincide under homoscedasticity. If $\tauu$ and $\taub$ are each homoscedastic, then it is straightforward to show that $\bspmmtwo$ and $\bspmle$ coincide. However, if $\tauu$ and $\taub$ are each homoscedastic,  $\bspmmone$ and $\bspure$ are not equal to each other, nor are they equal to $\bspmmtwo$ and $\bspmle$. 

\subsection{Confidence Intervals}\label{sec:CIs}

Valid confidence interval construction for shrinkage and data-combination estimators is an open area of research \citep{armstrong2020robust, hoff2019exact}. The results of \cite{chen2021minimax} indicate that frequentist confidence intervals for estimators combining observational and experimental data sources cannot be shortened, relative to those obtained from experimental data alone, when the magnitude of the confounding bias is unknown. For inference, we thus appeal to a common approach in the shrinkage literature, which is to use a relaxed notion of interval coverage known as ``Empirical Bayes (EB) coverage" \citep{armstrong2020robust, morris1983parametric}. 


Formally, EB coverage is the requirement that coverage holds over repeated resampling of the entire data-generating process, i.e. over both stages of the hierarchical model defined by Expression \ref{eq:obsData}. This is weaker than frequentist coverage, which would hold over repeated sampling of the data $\tauu$ and $\taub$ conditional on any value of the true causal effects $\bst$ and biases $\bsxi$. As discussed in  \cite{armstrong2020robust}, EB coverage also implies that, for any values of $\bst$ and $\bsxi$, the intervals should cover a $1 - \alpha$ proportion of the true values of $\bst$ across repeated sampling of the data. However, there may be some individual entries in $\bst$ which are covered in less than $ 1- \alpha$ fraction of samples of the data. 

We seek to obtain an interval construction procedure that provides valid EB coverage rates for all of $\bspmmone, \bspmmtwo, \bspmle,$ and $\bspure$. We pattern our construction exactly on the work of \cite{armstrong2020robust}, which constructed intervals for Stein-like estimators. Our goal is to design intervals that do not rely on the parametric assumptions about the distributions of $\bst$ and $\bsxi$, described in Equations \ref{eq:obsData}, in order to guarantee coverage. 

We suppose instead that all we know about $\bstau$ is that its entries are sampled from a distribution with a second moment $\eta^2$;
and all we know about $\bsxi$ is that its entries are sampled from a distribution with a second moment $\gamma^2$.
These second moments will be replaced by estimates in practice. In describing the construction procedure, we use the generic notation $\hat \eta^2$ and $\hat \gamma^2$ to refer to estimates of these hyperparameters. These quantities should be understood to represent, e.g. $\etammone$ and $\gammamm$ when using the estimator $\bspmmone$,  $\etamle$ and $\gammamle$ when using $\bspmle$, etc.

The full confidence interval derivation can be found in Appendix \ref{app:cis}. The resultant form for confidence interval for each stratum causal effect estimate $\psi_k$ is given in Definition \ref{def:cis}. 

\begin{definition}[Robust EB Confidence Intervals]\label{def:cis}
The robust EB confidence interval for $\psi_k$, the causal effect estimate for stratum $k$ obtained from any of our double-shrinkage estimators, is given by 
\[ \psi_k \pm cva(c_k) \hat a_k\sqrt{ \left( \hat \lambda_k^2 \sigma_{uk}^2 + (1 - \hat \lambda_k)^2 \sigma_{bk}^2 \right)}\,, \] 
where 
\[ \hat a_k = \left( 1 - \frac{\sigma_{uk}^2 \left(  \hat \gamma^2 + \sigma_{bk}^2 \right) }{\sigma_{uk}^2 \left( \hat \gamma^2 + \sigma_{bk}^2 \right) + \hat \eta^2 \left( \hat \gamma^2 + \sigma_{bk}^2 + \sigma_{uk}^2 \right) } \right), \hspace{5mm} \hat \lambda_k = \left( \frac{  \left(\hat  \gamma ^2+\sigma_{bk}^2\right) }{ \hat \gamma^2 + \sigma_{bk}^2 + \sigma_{uk}^2}\right), \]
and $cva(c_k)$ is a stratum-specific inflation factor whose precise form can be found in Appendix \ref{app:cis}.
\end{definition}

A final complicating detail that estimates of $ \eta^2$ and $ \gamma^2$ may be 0 in small samples, which can yield both the estimator and confidence intervals to concentrate on a single point, $\{0\}$. \cite{armstrong2020robust} considered this problem, and proposed an empirical truncation of the hyperparameter estimates to approximate the Bayesian posterior means under a flat prior -- an idea originated in \cite{morris1983parametric}. We apply the analogous truncation procedure in computation of our confidence intervals. For more details, see Appendix \ref{app:cis}.


\section{Simulations Using Data from the Women's Health Initiative}\label{sec:whi}

As an initial evaluation of our efficacy of our proposed estimators, we evaluate them on a familiar dataset: data from the Women's Health Initiative (WHI). The study is a 1991 study involving postmenopausal women, studying the health effects of hormone therapy. The WHI incorporated both a randomized controlled trial as well as an observational study. 

In total, 16,608 women were included in the trial. Half were randomly selected to take 625 mg of estrogen and 2.5 mg of progestin, while the remainder were given a placebo. The observational component of the WHI included 53,054 women clinically comparable to those in the trial. Roughly one third of women in the observational study used estrogen plus progestin. The remaining women were not using hormone therapy \citep{prentice2005combined}. 

\subsection{Risk Reduction}

As in \cite{rosenmanBiometrics}, we investigate the treatment's effect on coronary heart disease incidence. We draw $1,000$ bootstrap samples from the RCT component and observational component of the WHI. With the data from each bootstrap sample, we compute each of our double-shrinkage estimators as well as several competitor estimators. The bootstrap is a useful proxy for sampling from a super-population, as the causal estimates computed on the RCT bootstrap samples are normally distributed and centered on the ``true" causal estimates computed using the entire RCT sample. Thus, we can estimate the statistical risk of our estimators by computing their average mean squared error in estimating these true causal quantities. 

We use the same set of three stratification variables from \cite{rosenmanBiometrics}: two clinically relevant variables (cardiovascular disease history, or ``CVD", and age), as well as one clinically irrelevant variable (``Langley" scatter, a measure of solar irradiance). For full details on the choice of these variables, see  \cite{rosenmanBiometrics}. 

In Table \ref{tab:whiSims1000}, we provide results of $1,000$ simulations in which the pseudo-RCTs contain $1,000$ units. The rows correspond to different stratification schemes, created by stratifying on a subset of our stratification variables. In the second column, we give the number of strata. The following ten columns give the average mean square error of different estimators. For ease of interpretation, we report these average MSE values as a percentage of the average MSE of $\tauu$, which is a simple difference-in-means estimator applied to each stratum. 

Any value less than 100\% indicates a loss reduction. We compare our four proposed estimators -- $\bspmmone, \bspmmtwo, \bspmle$, and $\bspure$ -- against several competitors. We consider $\bsdelt_1$ and $\bsdelt_2$ from \cite{green2005improved}, as well as $\bskap_{1+}$ and $\bskap_{2+}$ from \cite{rosenmanBiometrics}. For reference we also consider $\taub$, the observational study estimator. We apply an inverse propensity weighting adjustment to $\taub$ using the propensity score generated in \cite{rosenman2020combining}, but we do not assume unconfoundedness holds in this setting, meaning residual bias remains in $\taub$. We also consider $\tauw$, a ``precision-weighted" estimator which computes a convex combination of $\tauu$ and $\taub$ where each estimator is weighted according to the inverse of its variance. 
In the given simulation regime -- in which the RCT is much smaller than the observational study, and a careful propensity score adjustment has been applied to the observational data -- both of these estimators have  lower MSE than the estimator computed from the small pseudo-RCTs, $\tauu$.

\begin{table}[h]
\centering
\begin{tabular}{llrrrrrrrrrr}
\toprule
&  & \multicolumn{9}{c}{Loss as a \% of $\tauu$ Loss}  \\ \cline{3-12} 
\begin{tabular}[c]{@{}l@{}}Subgroup \\ Vars\end{tabular}    & \begin{tabular}[c]{@{}l@{}}\# of \\ Strata\end{tabular} & $\htauo$ & $\tauw$ & $\bskap_{1+}$ & $\bskap_{2+}$ &$\bsdelt_1$ & $\bsdelt_2$ & $\bspmmone$ & $\bspmmtwo$ & $\bspmle$ & $\bspure$  \\ \midrule \\ \vspace{0.2cm}
CVD & 2 & 9\% & \underline{8\%} & 36\% & 36\% & 100\% & 100\% & 21\% & 17\% & 16\% & 32\% \\\vspace{0.2cm}
Age & 3 & 17\% & \underline{15\%} & 37\% & 30\% & 62\% & 73\% & 21\% & 18\% & 16\% & 34\% \\\vspace{0.2cm}
Langley & 5 & 23\% & 20\% & 28\% & 22\% & 39\% & 52\% & 11\% & 10\% & \underline{9\%} & 15\% \\\vspace{0.2cm}
CVD, Age & 6& 42\% & 36\% & 39\% & 42\% & 40\% & 83\% & 21\% & \underline{21\%} & 21\% & 27\% \\\vspace{0.2cm}
CVD, Langley & 10 & 35\% & 32\% & 34\% & 36\% & 33\% & 87\% & 17\% & \underline{17\%} & 17\% & 19\% \\\vspace{0.2cm}
Age, Langley & 15 & 21\% & 18\% & 22\% & 21\% & 21\% & 44\% & 8\% &\underline{7\%} & 8\% & 10\% \\\vspace{0.2cm}
\begin{tabular}[c]{@{}l@{}}CVD, Age,\\ Langley\end{tabular} & 30 & 51\% & 46\% & 51\% & 51\% & 51\% & 80\% & 20\% & \underline{19\%} & 20\% & 20\% \\ \bottomrule
\end{tabular}
\caption{\label{tab:whiSims1000} Simulation results for each stratification scheme, with an RCT sample size of $1,000$. The best-performing estimator is underlined for each stratification scheme.}
\end{table}

Results in Table \ref{tab:whiSims1000} are highly encouraging. In this data regime, the observational study estimator $\taub$ and the precision-weighted estimator $\tauw$ exhibit significantly lower MSE than the RCT-derived unbiased estimator $\tauu$. The four competitor shrinkers -- $\bskap_{1+}, \bskap_{2+}, \bsdelt_1$, and $\bsdelt_2$ -- are all able to achieve significantly lower risk than $\tauu$, but they typically do no better than $\taub$ and worse than $\tauw$. This is, at least in part, due to the fact that these estimators can only take data-driven convex combinations of $\tauu$ and $\taub$, and do not provide any additional stabilization via shrinkage toward zero.

The four ``double-shrinkers," $\bspmmone, \bspmmtwo, \bspmle,$ and $\bspure$, by contrast, all consistently outperform the competitor shrinkers $\bskap_{1+}, \bskap_{2+}, \bsdelt_1$, and $\bsdelt_2$. Moreover, we see that all four double-shrinkers are able to outperform $\tauw$ in five of the seven conditions, corresponding to all stratification schemes with five or more subroups. It appears that $\bspmle$ performs slightly better among stratifications with few subgroups, while $\bspmmtwo$ is especially performant when there are a larger number of subgroups. 
The performance of the double shrinkers, in aggregate, suggests that these estimators can do quite well in cases when the unbiased estimator has much higher variance than the biased estimator. 

Next, we consider a setting in which the RCT sample size is $8,000$ units. We provide these results in Table \ref{tab:whiSims8000}. The simulation set-up is otherwise identical. 

This setting is distinct because, due to the larger RCT sample size, $\taub$ has higher MSE than $\tauu$ in all stratifications involving three or more strata. As a consequence, the precision-weighted estimator $\tauw$ is also no longer particularly accurate, and is outperformed by the double shrinkers in all settings. The double shrinkers also consistently outperform each of $\bskap_{1+}, \bskap_{2+}, \delta_1$, and $\delta_2$. 

The best-performing double shrinker is less clear in this data regime. $\bspmle$ does best in four of the seven settings, while $\bspmmone$ does best in the remaining three. There is no clear relationship between the number of subgroups and the best-performing shrinker 

\begin{table}[h]
\centering
\begin{tabular}{llrrrrrrrrrr}
\toprule
&  & \multicolumn{9}{c}{Loss as a \% of $\tauu$ Loss}  \\ \cline{3-12} 
\begin{tabular}[c]{@{}l@{}}Subgroup \\ Vars\end{tabular}    & \begin{tabular}[c]{@{}l@{}}\# of \\ Strata\end{tabular} & $\htauo$ & $\tauw$ & $\bskap_{1+}$ & $\bskap_{2+}$ &$\bsdelt_1$ & $\bsdelt_2$ & $\bspmmone$ & $\bspmmtwo$ & $\bspmle$ & $\bspure$  \\ \midrule \\ \vspace{0.2cm}
CVD & 2 & 83\% & 49\% & 69\% & 62\% & 100\% & 100\% & 31\% & 29\% & \underline{25\%} & 47\% \\\vspace{0.2cm}
Age & 3 & 152\% & 76\% & 83\% & 95\% & 84\% & 90\% & 49\% & 48\% & \underline{44\%} & 65\% \\\vspace{0.2cm}
Langley & 5 & 173\% & 90\% & 77\% & 144\% & 77\% & 83\% & 29\% & 28\% & \underline{27\%} & 35\% \\\vspace{0.2cm}
CVD, Age & 6 & 233\% & 102\% & 87\% & 127\% & 80\% & 92\% & \underline{60\%} & 62\% & 64\% & 72\% \\\vspace{0.2cm}
CVD, Langley & 10 & 148\% & 66\% & 68\% & 103\% & 64\% & 95\% & \underline{33\%} & 34\% & 34\% & 43\% \\\vspace{0.2cm}
Age, Langley & 15 & 121\% & 66\% & 62\% & 104\% & 59\% & 81\% & 28\% & 29\% & \underline{28\%} & 37\% \\\vspace{0.2cm}
\begin{tabular}[c]{@{}l@{}}CVD, Age,\\ Langley\end{tabular} & 30 & 157\% & 67\% & 78\% & 147\% & 67\% & 95\% & \underline{38\%} & 39\% & 39\% & 41\% \\\bottomrule
\end{tabular}
\caption{\label{tab:whiSims8000} Simulation results for each stratification scheme, with an RCT sample size of $8,000$. The best-performing estimator is underlined for each stratification scheme.}
\end{table} 

We note briefly that $\bspure$ is a surprising laggard among the double shrinkers, rarely achieving the performance of $\bspmmone, \bspmmtwo$, and $\bspmle$. This is an unexpected result, as the SURE-minimizing estimator in \citep{xie2012sure} was consistently the best estimator. One hypothesis is that the URE is somewhat challenging to minimize precisely over the positive orthant, and it is feasible that the R function \texttt{optim()} -- used in our code -- can only achieve approximate optima. Another hypothesis is that the high-noise setting in which we simulate may be unattractive for use of URE minimization. 

Taken together, these simulation results demonstrate the practical improvement in point estimation that can be achieved using our new class of estimators. We next turn our attention to questions of inference. 

\subsection{Confidence Interval Coverage}

We also simulate confidence interval coverage, where the intervals are constructed using the robust Empirical Bayes method described in Section \ref{sec:CIs}. Again, we draw $1,000$ bootstrap samples from the data. In each sample, we construct the robust 95\% intervals for each of the double shrinkers, for each subgroup causal effect, under each of the seven stratifications. We then compute the frequency with which the intervals cover the ``true" causal effects computed from the entire RCT population, as well as their average lengths. 

We first consider the case when the RCT includes just $1,000$ units. In Table \ref{tab:whiSimCIs1000}, we report the average coverage rate across the subgroups under each stratification scheme. In Table \ref{tab:whiSimCILengths1000}, we provide the average confidence interval length; for ease of interpretation, we report the length as a percentage of the average length of a standard Wald interval computed using the RCT data only (i.e., the default confidence interval that would be used in the absence of the observational data). 

In Table \ref{tab:whiSimCIs1000}, we can see that the average coverage rate is consistently above the nominal rate of 95\%, indicating that Empirical Bayes coverage indeed holds in practice. In fact, we appear to somewhat overcover, a likely consequence of the robustness property of the intervals defined via Definition \ref{def:cis}. As discussed in Appendix \ref{app:cis}, these intervals are designed to achieve EB coverage against a ``least-favorable" distribution of $\bstau$ and $\bsxi$ which still satisfy a second-moment condition. If our distributions of $\bstau$ and $\bsxi$ are not the most adversarial distributions possible, we may have some overcoverage. 

Nonetheless, in Table \ref{tab:whiSimCILengths1000}, we observe that the intervals are consistently shorter than those computed the RCT data only, with length reductions ranging from 15 to 50\% and reductions typically larger in settings with more subgroups. Moreover, we see that there is very little variability across the four double shrinkage estimators. All the estimators consistently achieve the desired EB coverage rate, and all achieve similar length reductions. 

\begin{table}[h]
\centering
\begin{tabular}{llrrrrr}
\toprule
&  & \multicolumn{4}{c}{Coverage Rate}  \\ \cline{3-6} 
\begin{tabular}[c]{@{}l@{}}Subgroup \\ Vars\end{tabular}    & \begin{tabular}[c]{@{}l@{}}\# of \\ Strata\end{tabular} & $\bspmmone$ &  $\bspmmtwo$ & $\bspmle$ & $\bspure$  \\ \midrule \\ \vspace{0.2cm}
CVD & 2 & 98\% & 99\% & 99\% & 98\% \\\vspace{0.2cm}
Age & 3 & 100\% & 100\% & 100\% & 99\% \\\vspace{0.2cm}
Langley & 5 & 100\% & 100\% & 100\% & 100\% \\\vspace{0.2cm}
CVD, Age & 6 & 100\% & 100\% & 100\% & 100\% \\\vspace{0.2cm}
CVD, Langley & 10 & 99\% & 99\% & 99\% & 99\% \\\vspace{0.2cm}
Age, Langley & 15 & 100\% & 100\% & 100\% & 100\% \\\vspace{0.2cm}
\begin{tabular}[c]{@{}l@{}}CVD, Age,\\ Langley\end{tabular} & 30 & 100\% & 100\% & 100\% & 100\% \\ \bottomrule
\end{tabular}
\caption{\label{tab:whiSimCIs1000} Robust confidence interval average coverage rate for each stratification scheme, with an RCT sample size of $1,000$. The reported values are the average coverage levels, where averages are computed across the subgroups and over the $1,000$ bootstrap samples. }
\end{table}

\begin{table}[h]
\centering
\begin{tabular}{llrrrrr}
\toprule
&  & \multicolumn{4}{c}{\begin{tabular}[c]{@{}l@{}} Avg. Length as a \% of \\ RCT-Only Interval Length \end{tabular} }  \\ \cline{3-6} 
\begin{tabular}[c]{@{}l@{}}Subgroup \\ Vars\end{tabular}    & \begin{tabular}[c]{@{}l@{}}\# of \\ Strata\end{tabular} & $\bspmmone$ &  $\bspmmtwo$ & $\bspmle$ & $\bspure$  \\ \midrule \\ \vspace{0.2cm}
CVD & 2 &  83\% & 83\% & 83\% & 84\%  \\\vspace{0.2cm}
Age & 3 &  78\% & 78\% & 78\% & 79\%  \\\vspace{0.2cm}
Langley & 5 &  72\% & 72\% & 71\% & 72\%  \\\vspace{0.2cm}
CVD, Age & 6 &  70\% & 71\% & 70\% & 72\%  \\\vspace{0.2cm}
CVD, Langley & 10  & 66\% & 68\% & 66\% & 67\%  \\\vspace{0.2cm}
Age, Langley & 15  & 60\% & 63\% & 59\% & 61\%  \\\vspace{0.2cm}
\begin{tabular}[c]{@{}l@{}}CVD, Age,\\ Langley\end{tabular} & 30  & 50\% & 58\% & 50\% & 50\%  \\ \bottomrule
\end{tabular}
\caption{\label{tab:whiSimCILengths1000} Robust confidence interval average lengths, for each stratification scheme, with an RCT sample size of $1,000$. The reported values are scaled by the average length of a Wald interval computed using the RCT data alone, such that values less than $100\%$ indicate shorter intervals. }
\end{table}

In Tables \ref{tab:whiSimCIs8000} and \ref{tab:whiSimCILengths8000}, we provide the analogous results for the case when the RCT sample size is $8,000$. The story is essentially unchanged: Empirical Bayes coverage is achieved at the 95\% level, and the intervals are 15 to 50\% shorter on average than standard intervals computed using the RCT data alone. 

Targeting the weaker notion of Empirical Bayes coverage means that certain subgroups' causal effects can be undercovered by our procedure, as long as average coverage rates remain at or above the nominal level of $1 - \alpha$. This is discussed in more detail in Appendix \ref{sec:minRates}. 

\begin{table}[h]
\centering
\begin{tabular}{llrrrrr}
\toprule
&  & \multicolumn{4}{c}{Coverage Rate}  \\ \cline{3-6} 
\begin{tabular}[c]{@{}l@{}}Subgroup \\ Vars\end{tabular}    & \begin{tabular}[c]{@{}l@{}}\# of \\ Strata\end{tabular} & $\bspmmone$ &  $\bspmmtwo$ & $\bspmle$ & $\bspure$  \\ \midrule \\ \vspace{0.2cm}
CVD & 2  & 97\% & 97\% & 98\% & 95\% \\\vspace{0.2cm}
Age & 3  & 98\% & 98\% & 99\% & 97\% \\\vspace{0.2cm}
Langley & 5  & 98\% & 98\% & 99\% & 98\% \\\vspace{0.2cm}
CVD, Age & 6  & 97\% & 97\% & 97\% & 96\% \\\vspace{0.2cm}
CVD, Langley & 10 & 98\% & 98\% & 98\% & 97\% \\\vspace{0.2cm}
Age, Langley & 15 & 97\% & 97\% & 97\% & 97\% \\\vspace{0.2cm}
\begin{tabular}[c]{@{}l@{}}CVD, Age,\\ Langley\end{tabular} & 30 & 95\% & 95\% & 93\% & 93\% \\ \bottomrule
\end{tabular}
\caption{\label{tab:whiSimCIs8000} Robust confidence interval average coverage rate for each stratification scheme, with an RCT sample size of $8,000$. The reported values are the average coverage levels, where averages are computed across the subgroups and over the $1,000$ bootstrap samples. }
\end{table}

\begin{table}[h]
\centering
\begin{tabular}{llrrrrr}
\toprule
&  & \multicolumn{4}{c}{\begin{tabular}[c]{@{}l@{}} Avg. Length as a \% of \\ RCT-Only Interval Length \end{tabular} }  \\ \cline{3-6} 
\begin{tabular}[c]{@{}l@{}}Subgroup \\ Vars\end{tabular}    & \begin{tabular}[c]{@{}l@{}}\# of \\ Strata\end{tabular} & $\bspmmone$ &  $\bspmmtwo$ & $\bspmle$ & $\bspure$  \\ \midrule \\ \vspace{0.2cm}
CVD & 2 &  82\% & 82\% & 82\% & 83\% \\\vspace{0.2cm}
Age & 3 &  80\% & 80\% & 79\% & 82\% \\\vspace{0.2cm}
Langley & 5 & 74\% & 72\% & 72\% & 72\% \\\vspace{0.2cm}
CVD, Age & 6 & 76\% & 73\% & 71\% & 79\% \\\vspace{0.2cm}
CVD, Langley & 10  & 70\% & 68\% & 65\% & 70\% \\\vspace{0.2cm}
Age, Langley & 15  & 68\% & 67\% & 64\% & 70\% \\\vspace{0.2cm}
\begin{tabular}[c]{@{}l@{}}CVD, Age,\\ Langley\end{tabular} & 30  &56\% & 56\% & 51\% & 55\%  \\ \bottomrule
\end{tabular}
\caption{\label{tab:whiSimCILengths8000} Robust confidence interval average lengths, for each stratification scheme, with an RCT sample size of $8,000$. The reported values are scaled by the average length of a Wald interval computed using the RCT data alone, such that values less than $100\%$ indicate shorter intervals. }
\end{table}

\section{Discussion}\label{sec:conclusion}

Our work contributes to the active and growing literature on designing estimators to trade off between biased and unbiased estimators of causal effects \citep{oberst2022bias, chen2021minimax, yang2020elastic, cheng2021adaptive}. Building on ideas discussed in prior Empirical Bayes papers -- namely, those of \cite{green1991james} as well as \cite{xie2012sure} -- we propose two innovations. First, we obtain the functional form of our estimator by appealing to a hierarchical model and computing a posterior mean. This yields the novel structure of our shrinkage estimator, which empirically estimates a convex combination of the biased and unbiased estimators, and then also applies a stabilizing shrinkage toward zero. Second, we operationalize our estimator by considering several different paradigms for estimating hyperparameters in the hierarchical model. This yields four different versions of our ``double shrinkage" estimator, three of which -- $\bspmmone, \bspmmtwo,$ and $\bspmle$ -- appear to be consistently performant. 

We also propose a method for constructing robust confidence intervals to guarantee coverage under a weaker inferential paradigm known as ``Empirical Bayes coverage." This approach allows us to make use of the biased dataset in order to construct shorter confidence intervals for each subgroup causal effect. We demonstrate the utility of these methods using data from the Women's Health Initiative. Simulating from the data, we find that our double shrinkers consistently outperform competitor estimators in terms of mean squared error in estimating subgroup causal effects. Our confidence intervals also empirically achieve their nominal EB coverage rates, while achieving 15-50\% average reductions in interval length.

There are many potential areas for further work on this topic. We find very little difference between the performances of $\bspmmone, \bspmmtwo,$ and $\bspmle$ in our simulation study. Further work to identify the relative strengths of each estimator may yield stronger guidelines for when to make use of each estimator. 

Relatedly, the surprisingly poor performance of $\bspure$, the estimator built upon an unbiased risk estimate (URE), is one major contrast to the work of \cite{xie2012sure}. This presents an opportunity for further exploration and potential modification of the estimator. Two extensions discussed in \cite{xie2012sure} -- estimators that shrink toward a fixed point other than zero; and semiparametric shrinkers whose parameters are optimized over the URE -- represent potential future steps. 

The double-shrinkage paradigm can also be extended to at least three more complex cases. First, we might consider settings involving multiple observational studies and multiple experiments, in which case we would want to construct estimators to compute weights for each of $\boldsymbol{\hat \tau_{u1}}, \dots, \boldsymbol{\hat \tau_{un_u}}, \boldsymbol{\hat \tau_{b1}}, \dots, \boldsymbol{\hat \tau_{n_b}}$, and then shrink toward a fixed point. Second, we can consider the case beyond stratification, where CATEs are estimated by continuous functions of the covariates, $\tauu(x)$ and $\taub(x)$, which need to be combined and shrunken appropriately. Lastly, we can consider settings involving continuous treatments, such that we combine biased and unbiased estimators of dose-response curves rather than treatment effect estimates. 

\bibliographystyle{apalike}
\bibliography{rctodb}

\appendix

\section{Computation of Unbiased Risk Estimate}\label{app:ureComp}

In \cite{rosenmanBiometrics}, the authors consider the case of combining any unbiased estimator $\tauu \sim \dnorm(\bst, \bsSig_u)$ (where $\bsSig_u = \text{diag}(\sigma_{u1}^2,, \dots, \sigma_{uk}^2)$) with any biased estimator $\taub$. The authors show that for any shrinkage estimator of the form  
\[ \bskap(\tauu, \taub) =  \tauu +   g(\tauu, \taub), \] 
where $ g(\tauu,  \taub) $ is a function of $\tauu$ and $\taub$ that is differentiable and satisfies  $\E(|| g||^2) < \infty$, the expectation of the squared error loss of the shrinkage estimator is given by
\begin{equation}\label{eq:ureExp}
\E \left( || \bst - \bskap( \tauu,  Y) ||_2^2 \mid \bstau, \bsxi \right) =  \Tr\left( \bsSig  \right) + \E \left(\sum_{k = 1}^K   g_k^2(\tauu,  \taub) + 2 \sigma_{uk}^2 \frac{\partial g_k(\tauu,  \taub)}{\partial \tau_{uk}} \mid \bstau, \bsxi \right) \,. 
\end{equation}

A direct consequence of this result is that the expression 
\[ \Tr\left( \bsSig  \right) +  \left(\sum_{k = 1}^K   g_k^2(\tauu,  \taub) + 2 \sigma_{uk}^2 \frac{\partial g_k(\tauu,  \taub)}{\partial \tau_{uk}} \right) , \] 
obtained by removing the expectation from the righthand side of Equation \ref{eq:ureExp}, is an unbiased estimator of the risk of any shrinkage estimator $\bskap(\tauu, \taub)$ that lies in this class of estimators.  

Now, observe that we can write
\begin{align*}
\bsp(\hat \gamma^2, \hat \eta^2) &= \boldsymbol a \left( \hat \gamma^2, \hat \eta^2 \right) \circ  \bigg( \boldsymbol \lambda (\hat \gamma^2, \hat \eta^2)  \circ \tauu + \big(\boldsymbol 1 - \boldsymbol \lambda (\hat \gamma^2, \hat \eta^2) \big) \circ  \taub \bigg) \\
&= \tauu + \underbrace{\bigg(  \big(\boldsymbol 1 -  a \left( \hat \gamma^2, \hat \eta^2 \right) \circ  \boldsymbol \lambda (\hat \gamma^2, \hat \eta^2) \big) \circ \tauu + a \left( \hat \gamma^2, \hat \eta^2 \right) \circ \big(\boldsymbol 1 - \boldsymbol \lambda (\hat \gamma^2, \hat \eta^2) \big) \circ  \taub   \bigg)}_{g^{\star}(\tauu, \taub)},
\end{align*}
where the function $g^{\star}(\tauu, \taub)$ is indeed differentiable and square-integrable. Hence, $\bsp(\hat \gamma^2, \hat \eta^2)$ lies in the desired class of estimators, and the above results apply.

We observe that 
\[ \frac{\partial g^{\star}(\tauu, \taub)}{\partial \tau_{uk}} =  1 - a_k(\hat \gamma^2, \hat \eta^2) \lambda_k(\hat \gamma^2, \hat \eta^2)\,.  \] 
Thus, an unbiased estimate of the risk of the estimator is given by 
\[  \Tr ( \bsSig_u ) +  || g^{\star}(\tauu, \taub) ||_2^2   -  2\sum_k  \sigma_{uk}^2 \cdot \big(1 - a_k \left( \hat \gamma^2, \hat \eta^2 \right) \cdot \lambda_k \left( \hat \gamma^2, \hat \eta^2 \right) \big)   \]
which can be equivalently written as 
\[ \Tr ( \bsSig_u ) +  || \bsp(\hat \gamma^2, \hat \eta^2) - \tauu ||_2^2   -  2\sum_k  \sigma_{uk}^2 \cdot \big(1 - a_k \left( \hat \gamma^2, \hat \eta^2 \right) \cdot \lambda_k \left( \hat \gamma^2, \hat \eta^2 \right) \big) \,.  \]
We label this quantity $\ure(\hat \gamma^2, \hat \eta^2)$ to signify that it is an unbiased risk estimate which is a function of the values of the hyperparameters $\hat \gamma^2$ and $\hat \eta^2.$

\section{Confidence Interval Construction}\label{app:cis}

We use the generic notation $\bsp, \hat \eta, \hat \gamma, \hat a$, and $\hat \lambda$ to represent the estimator and parameter estimates that correspond to any of the moment-matching, maximum likelihood, or URE minimization procedures described in Section \ref{subsec:op}.

We work with the individual entries of $\bsp$, e.g.
\[ \psi_k = \hat a_k \left( \hat \lambda_k \hat \tau_{uk} + (1 - \hat \lambda_k) \hat \tau_{bk} \right) \] 
where 
\begin{equation}\label{eq:forms}
 \hat a_k = \left( 1 - \frac{\sigma_{uk}^2 \left(  \hat \gamma^2 + \sigma_{bk}^2 \right) }{\sigma_{uk}^2 \left( \hat \gamma^2 + \sigma_{bk}^2 \right) + \hat \eta^2 \left( \hat \gamma^2 + \sigma_{bk}^2 + \sigma_{uk}^2 \right) } \right) \text{ and } \hat \lambda_k = \left( \frac{  \left(\hat  \gamma ^2+\sigma_{bk}^2\right) }{ \hat \gamma^2 + \sigma_{bk}^2 + \sigma_{uk}^2}\right). 
\end{equation}

Our goal is to design robust intervals that do not rely on the normality assumptions for $\bst$ and $\bsxi$, described in our hierarchical model (Expression \ref{eq:obsData}). The normality of $\tauu$ and $\taub$ \emph{conditional} on $\tau$ and $\xi$ is still assumed, as it holds due to the Central Limit Theorem, given sufficient sample size and bounded potential outcomes. We proceed as though all we know about $\bstau$ is that its entries are sampled from a distribution with a second moment $\eta^2$; and all we know about $\bsxi$ is that its entries are sampled from a distribution with a second moment $\gamma^2$. Because the variances $\{\sigma_{uk}^2\}$ and $\sigma_{bk}^2$ are also assumed known, our derivation proceeds as though $a_k$ and $\lambda_k$ were fixed (as in \cite{armstrong2020robust}). In practice, we will replace $\eta^2$ with an estimate $\hat \eta^2$ and $\gamma^2$ with an estimate $\hat \gamma^2$. 

For fixed $a_k$ and $\lambda_k$, the conditional bias of $\psi_k$ is given by 
\begin{align*}
\E \left( \psi_k - \tau_k \mid \bstau, \bsxi \right) &= a_k \left( \lambda_k  \tau_{k} + (1 - \lambda_k) ( \tau_{k} + \xi_k) \right) - \tau_k \\
& = (a_k - 1) \tau_k + a_k (1 - \lambda_k) \xi_k 
\end{align*}
and the conditional variance is given by 
\[ \var \left( \psi_k - \tau_k \mid \bstau, \bsxi \right) = a_k^2 \left( \lambda_k^2 \sigma_{uk}^2 + (1 - \lambda_k)^2 \sigma_{bk}^2 \right)\,. \] 
For each entry $\psi_k$, we can construct a t-statistic as
\[ t_k = \frac{\psi_k - \tau_k}{\sqrt{a_k^2 \left( \lambda_k^2 \sigma_{uk}^2 + (1 - \lambda_k)^2 \sigma_{bk}^2 \right)}}, \] 
where we observe that 
\[ b_k \equiv \E(t_k \mid \bstau, \bsxi) = \frac{(1 - 1/a_k) \tau_k + (1 - \lambda_k) \xi_k }{\sqrt{\lambda_k^2 \sigma_{uk}^2 + (1 - \lambda_k)^2 \sigma_{bk}^2}}  \text{ and } \var(t_k \mid \bstau, \bsxi) = 1.\] 
It follows that for any choice of critical value $\chi$, the interval \\ $\psi_k \pm \chi a_k\sqrt{ \left( \lambda_k^2 \sigma_{uk}^2 + (1 - \lambda_k)^2 \sigma_{bk}^2 \right)}$ will have non-coverage probability
\[ r(b_k, \chi) = \P \left( | Z - b_k | \geq \chi \mid \bstau, \bsxi \right) = \Phi (- \chi - b_k) + \Phi(-\chi + b_k), \] 
where $Z$ denotes a standard normal random variable, and $\Phi$ denotes its CDF. 

Via the same argument of \cite{armstrong2020robust}, we now see that the worst-case non-coverage is bounded by 
\begin{equation}\label{eq:optProblem}
\begin{aligned}
     \rho(c, \chi) &= \sup_F \E_F( r(b_k, \chi)) \text{ s.t. }\\
    \E_F( b_k ^2) &= \E_F \left(\frac{(1 - 1/a_k)^2 \tau_k^2 + (1 - \lambda_k)^2 \xi_k^2 + 2(1 - 1/a_k)(1 - \lambda_k) \tau_k\xi_k}{\lambda_k^2 \sigma_{uk}^2 + (1 - \lambda_k)^2 \sigma_{bk}^2} \right) \\
    &= \frac{(1 - 1/a_k)^2 \E_F(\tau_k^2) + (1 - \lambda_k)^2\E_F( \xi_k^2) + 2(1 - 1/a_k)(1 - \lambda_k) \E_F(\tau_k\xi_k)}{\lambda_k^2 \sigma_{uk}^2 + (1 - \lambda_k)^2 \sigma_{bk}^2}\\
    &= \frac{(1 - 1/a_k)^2  \eta^2 + (1 - \lambda_k)^2 \gamma^2 }{\lambda_k^2 \sigma_{uk}^2 + (1 - \lambda_k)^2 \sigma_{bk}^2} \\
    &= \frac{\sigma_{uk}^2 \left(\gamma ^2 \left(\eta ^2+2
   \sigma_{bk}^2\right)+\gamma ^4+\sigma_{bk}^4\right)}{\eta ^2
   \left(\left(\gamma ^2+\sigma_{bk}^2\right)^2+\sigma_{bk}^2
   \sigma_{uk}^2\right)}
\end{aligned}
\end{equation}

Here, $\E_F(\cdot)$ denotes the expectation associated with the distribution $F$ where $b_k \sim F$. For simplicity of notation, we refer to the quantity given in the final line as $c_k$, i.e. 
\[ c_k \equiv  \frac{\sigma_{uk}^2 \left(\gamma ^2 \left(\eta ^2+2
   \sigma_{bk}^2\right)+\gamma ^4+\sigma_{bk}^4\right)}{\eta ^2
   \left(\left(\gamma ^2+\sigma_{bk}^2\right)^2+\sigma_{bk}^2
   \sigma_{uk}^2\right)}\] 
In words, we're finding the worst-case non-coverage probability that is possible for any distribution of $\bstau$ and $\bsxi$ which still satisfies the constraints on the second moments of $\bst$ and $\bsxi$. 

Fortunately, \cite{armstrong2020robust} find a simple, closed-form solution to Problem \ref{eq:optProblem}. This is especially useful, because it means we can invert the procedure to generate robust confidence intervals by choosing a sufficiently large value of $\chi$. In particular, we set $\chi = cva(c_k)$ where $cva(t) = \rho^{-1}(t, \alpha)$ (and the inverse is with respect to the second argument). If we then define our confidence interval as 
\[ \psi_k \pm cva(c_k) a_k\sqrt{ \left( \lambda_k^2 \sigma_{uk}^2 + (1 - \lambda_k)^2 \sigma_{bk}^2 \right)}\,, \] 
we will maintain coverage under all distributions consistent with our second moment constraints. 

\section{WHI Simulations: Minimum Coverage Rates}\label{sec:minRates}

Tables \ref{tab:whiSimCIs1000} and \ref{tab:whiSimCIs8000} provide the average coverage rate for the confidence intervals generated when sampling pseudo-RCTs of size $1,000$ and $8,000$. In Tables \ref{tab:whiSimCIs1000Min} and \ref{tab:whiSimCIs8000Min}, we provide an alternate statistic: the \emph{minimum} coverage rate, across the subgroups, in each stratification. A key distinction between the frequentist coverage paradigm and the Empirical Bayes coverage paradigm is that the latter allows individual subgroup causal effects to be covered with less than $1 - \alpha$ probability as long as the average coverage rate across subgroups is at least $1 - \alpha$. We see precisely this behavior in Table \ref{tab:whiSimCIs8000Min} in some of the finer stratifications: certain subgroups causal effects are under-covered, such that the minimum coverage rate is well below $1 - \alpha$. This is the ``price" we pay, in some sense, for the shorter intervals achieved by our Empirical Bayes procedure. 

\begin{table}[h]
\centering
\begin{tabular}{llrrrrr}
\toprule
&  & \multicolumn{4}{c}{Coverage Rate}  \\ \cline{3-6} 
\begin{tabular}[c]{@{}l@{}}Subgroup \\ Vars\end{tabular}    & \begin{tabular}[c]{@{}l@{}}\# of \\ Strata\end{tabular} & $\bspmmone$ &  $\bspmmtwo$ & $\bspmle$ & $\bspure$  \\ \midrule \\ \vspace{0.2cm}
CVD & 2  & 97\% & 97\% & 98\% & 95\% \\\vspace{0.2cm}
Age & 3  & 98\% & 98\% & 99\% & 97\% \\\vspace{0.2cm}
Langley & 5  & 98\% & 98\% & 99\% & 98\% \\\vspace{0.2cm}
CVD, Age & 6  & 97\% & 97\% & 97\% & 96\% \\\vspace{0.2cm}
CVD, Langley & 10 & 98\% & 98\% & 98\% & 97\% \\\vspace{0.2cm}
Age, Langley & 15 & 97\% & 97\% & 97\% & 97\% \\\vspace{0.2cm}
\begin{tabular}[c]{@{}l@{}}CVD, Age,\\ Langley\end{tabular} & 30 & 95\% & 95\% & 93\% & 93\% \\ \bottomrule
\end{tabular}
\caption{\label{tab:whiSimCIs1000Min} Robust confidence interval \emph{minimum} coverage rate for each stratification scheme, with an RCT sample size of $1,000$. The reported values are the minimum coverage levels across the subgroups in each stratification. }
\end{table}

\begin{table}[h]
\centering
\begin{tabular}{llrrrrr}
\toprule
&  & \multicolumn{4}{c}{Coverage Rate}  \\ \cline{3-6} 
\begin{tabular}[c]{@{}l@{}}Subgroup \\ Vars\end{tabular}    & \begin{tabular}[c]{@{}l@{}}\# of \\ Strata\end{tabular} & $\bspmmone$ &  $\bspmmtwo$ & $\bspmle$ & $\bspure$  \\ \midrule \\ \vspace{0.2cm}
CVD & 2  & 95\% & 95\% & 96\% & 94\% \\\vspace{0.2cm}
Age & 3  & 97\% & 97\% & 97\% & 96\% \\\vspace{0.2cm}
Langley & 5  & 97\% & 96\% & 97\% & 96\% \\\vspace{0.2cm}
CVD, Age & 6  & 93\% & 92\% & 92\% & 92\% \\\vspace{0.2cm}
CVD, Langley & 10 & 93\% & 93\% & 92\% & 91\% \\\vspace{0.2cm}
Age, Langley & 15 & 85\% & 80\% & 71\% & 80\% \\\vspace{0.2cm}
\begin{tabular}[c]{@{}l@{}}CVD, Age,\\ Langley\end{tabular} & 30 & 44\% & 48\% & 9\% & 25\% \\ \bottomrule
\end{tabular}
\caption{\label{tab:whiSimCIs8000Min} Robust confidence interval \emph{minimum} coverage rate for each stratification scheme, with an RCT sample size of $8,000$. The reported values are the minimum coverage levels across the subgroups in each stratification. }
\end{table}

\end{document}